\documentclass[12pt]{iopart}
\expandafter\let\csname equation*\endcsname\relax
\expandafter\let\csname endequation*\endcsname\relax
\usepackage{amsmath}
\usepackage{gensymb}
\usepackage{graphicx}
\usepackage{epstopdf}
\usepackage{subfig}
\usepackage[numbers]{natbib}
\usepackage{tabularx}
\usepackage{miller}
\begin{document}
\title{An \textit{Ab Initio} Study of Aluminium self-compensation in Bulk Silicon}
\author{J.Poulton$^{1,2}$\footnote{corresponding author} and D.R. Bowler$^{1,2,3}$}
\address{$^1$ London Centre for Nanotechnology, 17-19 Gordon St, London, WC1H 0AH, UK}
\address{$^2$ Department of Physics \& Astronomy, UCL Gower St, London, WC1E 6BT, UK}
\address{$^3$ International Centre for Materials Nanoarchitectonics (MANA), National Institute for Materials Science
1-1 Namiki, Tsukuba, Ibaraki, 305-0044, Japan}
\ead{jack.poulton.15@ucl.ac.uk}
\begin{abstract}
\\
We have used density functional theory to study the energetics and electronic structure of aluminium dopants in crystalline silicon.  We present data regarding the atomic and electronic structure and properties of pairs of substitutional aluminium dopants. We find that  pairs of  dopants, when occupying nearest neighbouring subsitutional sites in a high spin state, can bond to form aluminium pairs. This suggests that such a configuration of dopants will be electrically active when made to occupy a high spin state, whereas in the low spin state the neighbouring dopant pairs are found to be  self compensating. 
\end{abstract}

\submitto{\JPCM}
\maketitle

\section{Introduction}

\subsection{Background}

The process of doping has proved a crucial tool for manipulating the physical properties of semiconductors and has been in widespread use for decades following the development of the first transistor in the mid $20^{th}$ century. Although a relatively low dopant concentration ($≈10^{13}$ atoms cm$^{-3}$) is sufficient to alter the material properties of a substrate, the decreasing dimensions of modern devices has required a corresponding increase in dopant concentration\citep{Dennard1999}. \\
Silicon technology is approaching scales where device characteristics are determined not only by the dopant concentration but also by the location of individual dopant atoms. It is expected that variability in device performance will be caused by the statistical nature of dopant placement\citep{Shinada2005}, which will impose limits in scalability prior to the physical limits associated with lithography and quantum effects. Hence precision doping is likely to become an essential tool. The need for precision has resulted in the development of patterned atomic layer epitaxy (PALE), which uses STM based lithography of H-passivated Si(001), and has been utilised to fabricate a FET transistor based on a single P atom using phosphine as precursor\citep{Fuechsle2012,Lyding1994,Owen2011,Shen1995}. \\
This progress has not extended to acceptor dopants. Although boron is often used as a p-type dopant, diborane is not a suitable candidate for a precursor in the PALE process due to the inability to selectively deposit single boron atoms\citep{Wang1995,Wang1996}. Furthermore boron would induce strain when used for delta-doping due to its small size relative to Si, resulting in fast diffusion rates\cite{Sarubbi2010} which would not give the precise dopant profile desired.  However the amine alanes are known to be viable precursors for the thin film deposition of aluminium\citep{Jones2008}, therefore Al may be a suitable candidate for the precision doping of Si. This precision doping could allow for fully compensated  co-doping, complementing P doping, which can be used to effectively tune the dopant populations, electronic properties, and magnetic properties\citep{Zhang2016}.  Alternatively, pairs of Al may self-compensate, as doped semiconductors can form defects or complexes that result in a gain overall free energy\citep{Mandel1964,Longini1956} albeit offset by the energy required to form the defect. \\ 
This work is motivated by the expectation that Al will become a viable candidate for precision doping in Si\cite{Smith2017}. There is therefore a need to determine whether a self-compensated effect could occur in highly doped samples where Al dopants occupy neighbouring sites but form a complex where the dopants are electrically inactive. 
The aim of this study is to test, with the use of \textit{ab initio} calculations, that adjacent Al dopants do not self compensate and therefore are electrically active.

\section{Methods}

\subsection{Computational Details}

All calculations were performed using density functional theory, as implemented in version 5.4.1 of the Vienna ab initio simulation package (VASP) software\citep{Joubert1999}. The exchange correlation potential has been approximated by the Perdew-Burke-Enzerhof (PBE) generalised gradient approximation (GGA)\citep{Perdew1996} functional. The VASP projector-augmented-wave (PAW)\citep{Blochl1994} potentials for Al and Si were used with  a 250~eV energy cut-off, which will incorporate all plane waves from the basis sets and allow consistency across all calculations. These potentials describe both core and valence electrons and the files (POTCAR) were dated 4/5 January 2001 and 15th June 2001, respectively.\\ All lattice parameters and atomic positions have been simultaneously optimised using a quasi-Newton relaxation algorithm giving near zero stress and near zero forces.  The convergence criteria  for the atomic forces and for the electronic structure were set to  0.01  eV/\AA\   to  $1 \times 10^{-6}$ eV respectively. These parameters yield relative energies reliable to within 0.02 eV when the Brillouin zone sampling mesh is set appropriately.  The Monkhorst-Pack\citep{Monkhorst1976} k-point sampling employed in these calculations for the 64, 216 and 512 atom supercells detailed below, were  ($4 \times 4 \times 4$),($3 \times  3 \times 3$) and ($2 \times 2 \times 2$) respectively. The energy values were found to converge for these Brillouin zone samplings and give consistent sampling of the Brillouin zone.\\
For analysis and the production of all two-dimensional charge density and electron localisation function plots shown in this work the visualisation software VESTA\citep{Momma2011} was used.

\subsection{Structural Models}

In this study cubic supercells of tetrahedral bulk silicon are constructed  and are subject to periodic boundary conditions.  The experimental bulk Si lattice parameter of $5.431$ \AA \citep{Becker1994}  was used as a starting value for all structures considered prior to the  relaxation of lattice parameters and atomic positions.  The aluminium dopants were placed at substitutional sites within the silicon structures with the total number of atoms modelled ranging from 64 to 512 atoms. The different cell sizes are used to explore the effect of system size on dopants, to adjust the isolation of the dopants and determine whether there is a non-negligible influence of the periodic images of the dopants which could affect bonding. For the electronic structure, a single reliable cell size will be used. Both single Al dopants and pairs of Al dopants are considered, with the pairs occupying either adjacent or non-adjacent substitutional sites. The different ratios of Si and Al atoms will be denoted as Si$_{n}$Al$_{m}$ throughout, with $n$ and $m$ representing the number of silicon and aluminium atoms respectively.

\subsection{Electronic Structure}

\subsubsection{Density of States}\hspace*{\fill} \\

In order to investigate any self compensation of the Al-Al pairing and the localisation of the associated holes the projected density of states (PDOS) will be considered.  The PDOS is produced by projecting the orbitals onto spherical harmonics which have non-zero values within spheres around each atom, these spheres have radii equal to the Wigner-Seitz radius\cite{NeilWAshcroft1976} of each species. The partial occupancies for the PDOS are determined using  Gaussian smearing techniques, using a smearing width of $0.02~$eV.

\subsubsection{Electron Localisation Function}\hspace*{\fill} \\

In order to investigate the presence of bonds between Al dopants, we have used the electron localisation function (ELF)\citep{Becke1990}, which is a function of the  spatial coordinates that has a high value in areas of concentrated electron density and can give a useful quantitative representation of chemical bonds\cite{Savin1997}. It is used here to investigate the presence of bonding between Al dopants. The function is dependent on the filled orbitals and density $\rho$, it is  defined as follows:
\begin{align}
\eta(r) = \frac{1}{1+(\frac{D}{Dh})^2}, 
\end{align}
\begin{align}
D = \frac{1}{2}\sum_{i=1}^{N}\nabla|\psi_{i}|^{2}-\frac{1}{8}\frac{| \nabla\rho|^{2}}{\rho},
\end{align} 
where $D(\textbf{r})$ is the probability of an electron being in the proximity of an electron of the same spin. For a homogeneous electron gas this probability has the value $D_{h}$:
\begin{align}
D_{h}=\frac{3}{10}(3\pi^{2})^{\frac{2}{3}}\rho^\frac{5}{3},
\end{align}
here the standard definition of electron density has been used where $\rho$ is:
\begin{align}
\rho = \sum_{i=1}^{N}|\psi_{i}|^2,
\end{align}
for the Kohn-Sham orbitals $\psi_{i}(\textbf{r})$.  As shown above, the ELF is inversely proportional to $D(\textbf{r})$. Therefore a low probability results in a high ELF which represents a localised pair of electrons. The ELF values are bound by 0 and 1 so a perfectly localised orbital has a value of 1, a  homogenous electron gas will have a value of 0.5 and  values will tend to 0 as electrons become completely uncorrelated. 

\section{Results and Discussion}

\subsection{Structural Features}

The simulations containing a single substitutional Al dopant have identical structural features with a constant Si-Al bond length of $2.43$~\AA\ for all cell sizes considered for both the $S=\frac{1}{2}$ and $S=0$ spin states.  The $S=0$ state has been modelled for comparison with the spin free pairs, and to represent metallic silicon. For both spin states this indicates an increase in bond length, following relaxation, around the dopant atom from the equilibrium Si-Si separation of $2.37$~\AA\ in pure Si. This is consistent with results which have been previously reported in the literature\cite{Ramos2007}. \\
In the cases where pairs of Al dopants are initially positioned in adjacent substitutional sites, there is an expansion of the Al-Al separation following ionic relaxation for both the $S=1$ and $S=0$ spin states. The magnitude of this expansion is dependent on whether spin polarisation is included in the simulation despite there being an even number of electrons present. When spin is introduced and the holes associated with the dopants are both forced to occupy either a spin up or spin down state then the Al-Al separation is comparatively smaller than the case where there is no net spin, the complete set of separation values is shown in Table 1. In both cases, the dopant separations are greater than the bulk silicon separations of $2.37$\AA\ found in  these simulations.\\

\begin{table}[h]
\caption{\label{label}The separations between adjacent Al dopants and the nearest neighbouring Si atoms, shown as Si$_{NN}$ with the bond angles and relaxed lattice constants for spin polarised and spin free simulations.}
\resizebox{\columnwidth}{!}{%
\begin{tabular}{| c | c | c | c | c | c | c |}
\hline
\textbf{Structure} &\textbf{Net Spin} &\textbf{ Al-Al Separation(\AA)} & \textbf{Al-Si\textsubscript{NN}(\AA)} & \textbf{\angle Al-Si\textsubscript{NN}(\degree)} & \textbf{Si\textsubscript{NNN}-Si(\AA)} & \textbf{Lattice  Constant(\AA)}\\
\hline
Si$_{62}$Al$_{2}$ (adjacent dopant)& 1 &  2.61 &  2.43 &  112.71 & 2.35  &5.4807\\
& 0 & 2.85 & 2.42 & 115.82 & 2.36 &  5.4809 \\
\hline
Si$_{214}$Al$_{2}$ (adjacent dopant) & 1 & 2.62 &  2.42 & 112.86 & 2.35 & 5.4695  \\
& 0 & 2.84 & 2.41 & 115.64 & 2.35 &   5.4696 \\
 \hline
Si$_{510}$Al$_{2}$ (adjacent dopant)& 1 & 2.62 & 2.42 & 112.84 & 2.35 & 5.4668 \\
& 0 & 2.83 & 2.41 & 115.60 & 2.36 & 5.4668 \\
 \hline
\end{tabular}%
}
\end{table}

 The structure of the cell has converged for the cell size of 216 atoms, as shown by both the Al-Al separations and the lattice constants. The  separations for the $S=1$ case are in good agreement with Al-Al bond lengths in small aluminium clusters previously obtained from DFT simulations\citep{Ojwang2008} as well as the experimental Al$_{2}$ dimer bond length of $2.70~$\AA\citep{Fu2002}. The non-spin-polarised separations show closer agreement to values obtained for bond lengths within bulk aluminium cells of $2.86~$\AA\citep{Jacobs2002}. 
	This disparity can not be attributed to a deformation of the cell size with the introduction of spin, as both the spin polarised simulations and those without spin polarisation have similar lattice constants following cell relaxation. The  tetrahedral symmetry is  preserved beyond the next nearest neighbours of the dopants in simulations with either spin configuration. However, the local tetrahedral symmetry is broken, as shown by the bond angles in Table 1, with the $S=1$ case showing a greater distortion from the  tetrahedral angle of $109.28\degree$. This is as a result of Si-Al bond lengths that are marginally smaller than the bond lengths observed in pure Si bulk.  These bonds lengths are also comparable to those obtained for the simulations containing single Al dopants, measured at $2.43$ \AA . Hence the differences between the two spin cases suggest there  is  a bond present between the Al-Al dopants for the $S=1$ case, a possibility that we will go on to investigate.
\

\subsection{Electronic Structure}

\subsubsection{Single Substitutional Aluminium Dopants}\hspace*{\fill} \\


The presence of Al-Si bonding is investigated by observation of the total charge density, as shown for the dopant plane in Figure 1.  The 216 atom structures have been used to produce all  plots containing a single atomic plane as the structures have converged at this cell size.  The charge density in the regions between the Al dopant and the neighbouring Si atoms is measured at approximately  $95\%$ of the magnitude of the charge density in the regions occupied by Si-Si bonds. This indicates that the Al will have bonded to its neighbouring Si atoms forming an unoccupied dopant state hence the preservation of tetrahedral symmetry. The regions of such high charge density are consistent with electron localisation and therefore further analysis of bond with the ELF is unnecessary. \\

\begin{figure}[h!]
\vspace{-0.4cm}
\centering
\hspace{-2mm}
\subfloat[Si$_{215}$Al$_{1}$ S=0]{
  \includegraphics[scale=0.40]{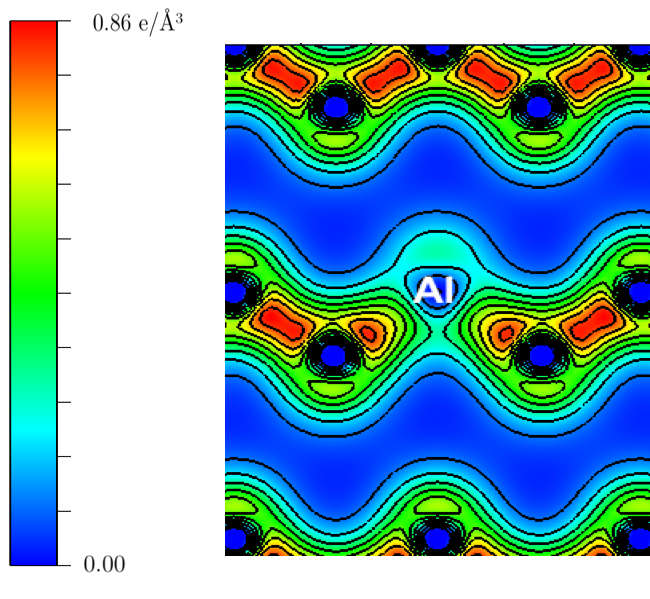}
}
\subfloat[Si$_{215}$Al$_{1}$  $S=\frac{1}{2}$]{
  \includegraphics[scale=0.40]{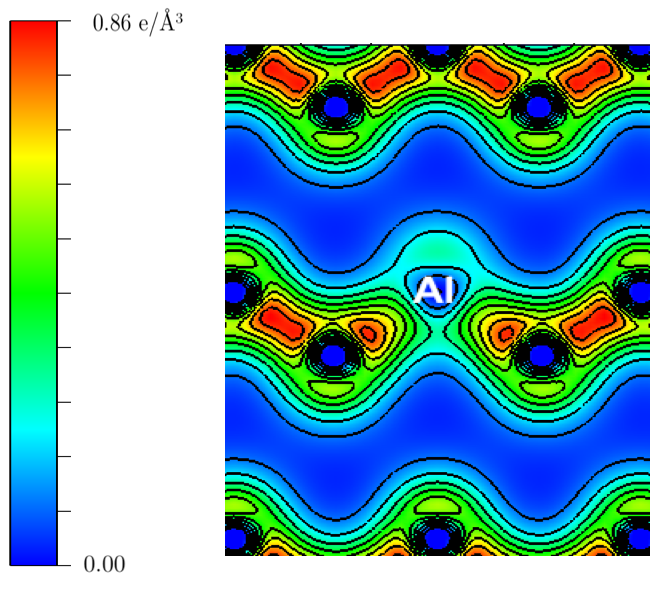}
}
\caption[]{A 2D plot of total charge density across the $\hkl{1,0,-1}$ plane containing the dopant for the Si$_{215}$Al$_{1}$ structure in the $S=\frac{1}{2}$  and $S=0$ spin states}
\end{figure}

From the analysis of the density of states, it can be seen that the incorporation of the Al dopant results in the presence of a shallow acceptor level in the gap  for both the $S=\frac{1}{2}$ and $S=0$ simulations as shown in  Figures 2 and 3 respectively. The bandgap is equal for both the $S=\frac{1}{2}$ and $S=0$ states.  The atom PDOS shows that the introduction of the Al dopant has resulted in a shift of the Fermi level into the VBM, as is typical for a p-type dopant, furthermore much clearer peaks can be observed for  $S=\frac{1}{2}$ case than in the $S=0$ case, suggesting a more localised acceptor state in the former.


\begin{figure}[h!]
\begin{center}
\hspace{-1cm}
\includegraphics[scale=0.55]{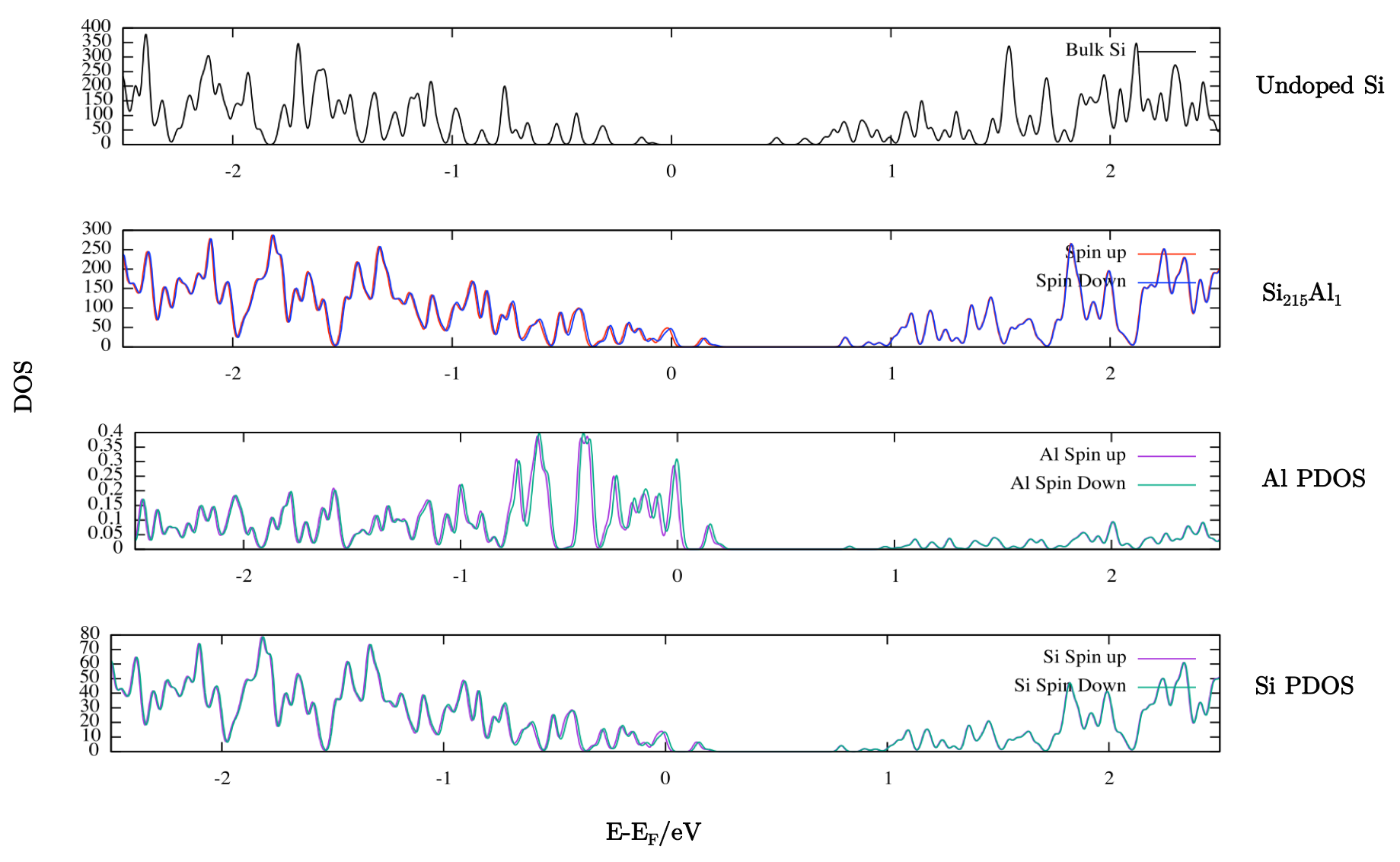}
\caption[]{The total and atom projected DOS for both undoped Si and the Si$_{215}$Al$_{1}$ structure in the $S=\frac{1}{2}$ spin state.}
\end{center}
\end{figure}

\begin{figure}[h!]
\begin{center}
\hspace{1cm}
\includegraphics[scale=0.55]{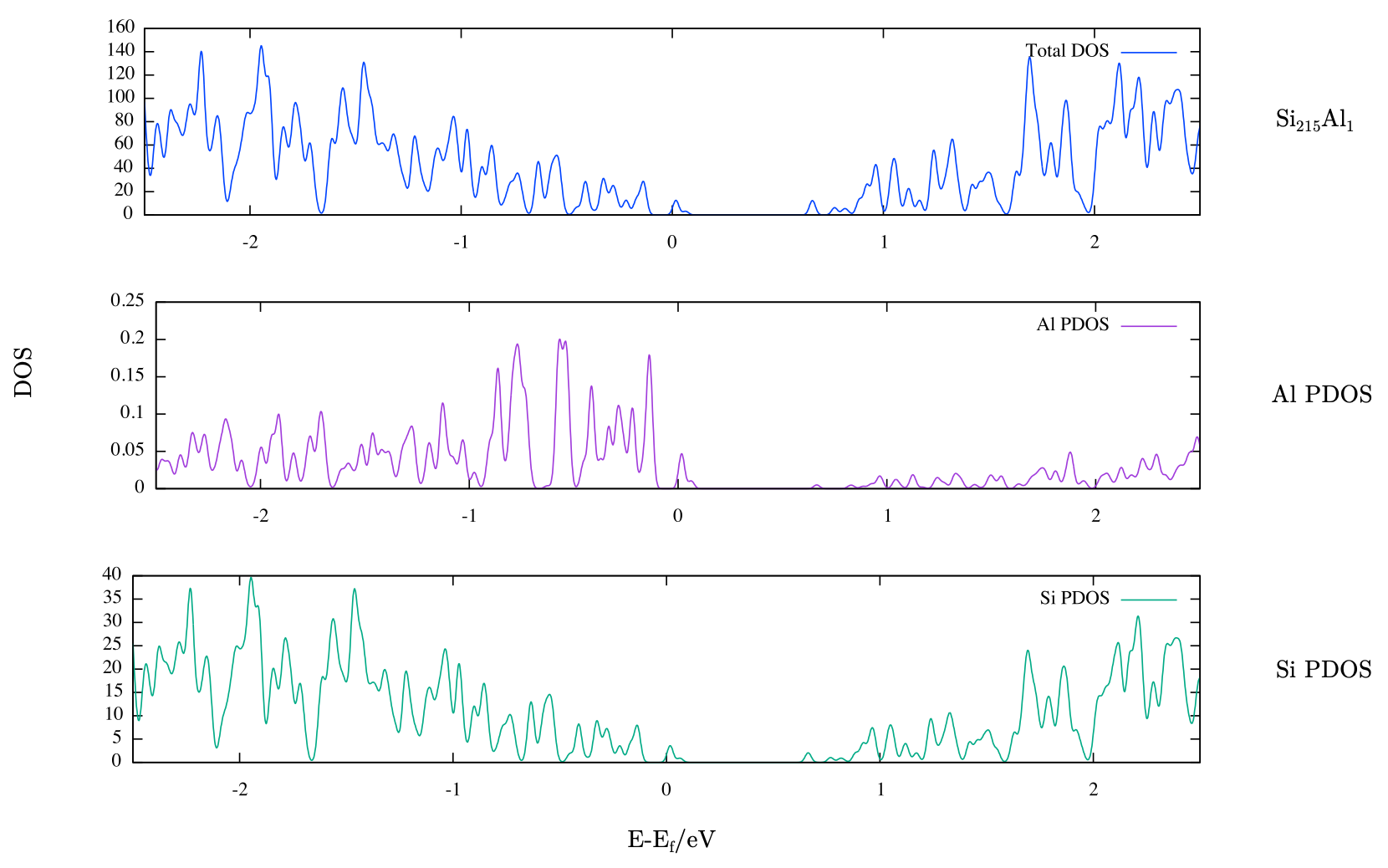}
\caption[]{The total and atom projected DOS for the Si$_{215}$Al$_{1}$ structure in the $S=0$ spin state. The total DOS is shown in the top panel with the  Al and Si pdos in the middle and bottom panels respectively.}
\end{center}
\end{figure}
\subsubsection{Substitutional Aluminium Pairs}\hspace*{\fill} \\


The single dopant Si$_{215}$Al$_{1}$ model is representative of a dopant concentration of $2.3\!~\times~\!10^{20}$cm$^{-3}$, this  beyond the metal-insulator transition for both boron\citep{Dai1992} and phosphorous\citep{FerreiraDaSilva1988} both of the order of  $1\!~\times~\!\!10^{18}$cm$^{-3}$.  At dopant concentrations of this order it is probable that a sample will contain dopants at nearby and even adjacent sites. Similarly Al could be unintentionally doped at adjacent sites as a consequence of the efficacy of the PALE process. This possible proximity could  result in the dopants self-compensating, to determine whether such an effect is present we consider simulations of adjacent and non-adjacent substitutional Al pairs. \\
 The relative energetic stability of doped structures, containing  both pairs of dopants and single dopants, can be determined by their formation energies as given by equation 5,

\begin{equation}
E_{defect}^{f}=\frac{E_{doped}-\left( \frac{N_{Si_{doped}}}{N_{Si_{bulk}}}\right) *E_{bulk}}{N_{Al}}
\end{equation} 

where $E_{bulk}$ is the energy of the optimised bulk undoped cell, $E_{doped}$ is the energy of the optimised doped cell and $N_{Si_{doped}}$ and $N_{Si_{bulk}}$ are the total numbers of Si atoms in each type of structure. $N_{Al}$ is simply the number of substitutional aluminium dopants  introduced to a given doped structure. 

The  formation energies shown in Table 2  demonstrate that for larger systems the substitutional impurities have greater stability, likely due to the reduced interaction with the periodic images of the dopants. It is interesting to note that there is variation in the formation energies of pairs of dopants at adjacent and non-adjacent sites within the two largest supercells considered here.  There is a small reduction in stability associated with the dopants occupying adjacent substitutional sites for the $S=\frac{1}{2}$ however this is not necessarily prohibitive to the formation of such a structure, as there is a small driving force encouraging the Al atoms to form pairs.  \\
\begin{table}[h]
\caption{\label{label}The formation energies of the different spin polarised  and spin free structures with pairs of dopants occupying separated and  adjacent substitutional sites labelled as `sep' and  `adj' respectively}
\begin{indented} 
\item[]\begin{tabular}{|c| c| c |}
\hline
\textbf{Structure} & $\mathbf{E_{defect}^{f}}$  $\mathbf{S\neq0}$ \textbf{(eV)} & $\mathbf{E_{defect}^{f}}$  $\mathbf{S= 0}$ \textbf{(eV)} \\
\hline
Si$_{63}$Al$_{1}$ & -2.71  & -2.74\\
 \hline
Si$_{62}$Al$_{2}$ (sep) & -2.67 & -2.76 \\
 \hline
Si$_{62}$Al$_{2}$ (adj) & -2.67 & -2.84 \\
 \hline
Si$_{215}$Al$_{1}$ & -2.73 & -2.77 \\
 \hline
Si$_{214}$Al$_{2}$ (sep) & -2.73 & -2.78 \\
 \hline
Si$_{214}$Al$_{2}$ (adj) & -2.69  & -2.81 \\
 \hline
Si$_{511}$Al$_{1}$ & -2.73 & -2.73 \\
 \hline
Si$_{510}$Al$_{2}$ (sep) & -2.76 & -2.79 \\
 \hline
Si$_{510}$Al$_{2}$ (adj)& -2.69 & -2.80  \\
 \hline
\end{tabular}
\end{indented}
\end{table}

The formation energies are  spin dependent with the non-spin-polarised structures being energetically favourable for both the single Al dopant and the  pairs of Al dopants.  It is also evident that the formation 
energy per Al  dopant is lower when pairs of dopants are introduced rather than a single Al dopant, this is due to the lack of partial occupancy in the multiple dopant case.



In addition to the differences in separations between the adjacent Al dopants in the two possible spin states, a change in the electronic structure is observed. As shown in Figure 4 the charge density between adjacent dopants in the $S=1$ spin state is of a greater magnitude than in the same region for the $S=0$ state. For the $S=1$ state the charge density between the Al dopants has a value of approximately $60\%$ of the density in the region between Si-Si bonds, opposed to a value of $41\%$ of the Si-Si bond density for the $S=0$ case, indicating that a bond is likely to be present for the $S=1$ case but not in $S=0$ simulation. 

\begin{figure}[h!]
\vspace{-0.4cm}
\centering
\hspace{-2mm}
\subfloat[Si$_{214}$Al$_{2}$ S=0]{
  \includegraphics[width=48mm]{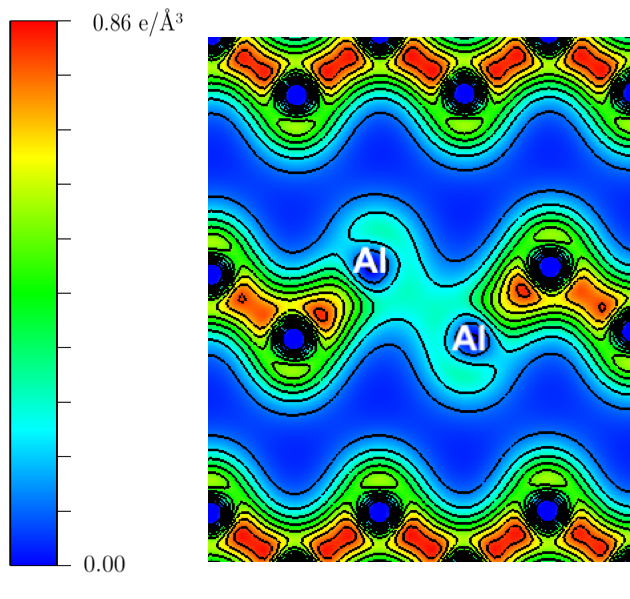}
}
\subfloat[Si$_{214}$Al$_{2}$  S=1]{
  \includegraphics[width=50mm]{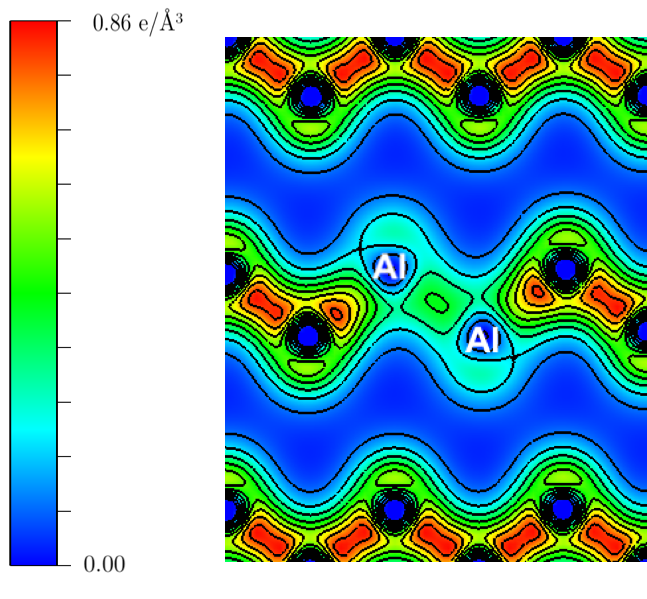}
}
\caption[]{A 2D plot of total charge density across the  $\hkl{1,0,-1}$ plane containing the dopant for the Si$_{214}$Al$_{2}$ structure in the $S=1$  and $S=0$ spin states}
\end{figure}


The differences in the electronic structure of the two spin states are also reflected in the DOS, as shown in Figures 5 and 6. The adjacent dopant structure in both spin states exhibits the creation of shallow acceptor states, as is the case for the single Al dopant. However the Al PDOS in the $S=1$ state, which is the same both dopants, shows a  spin splitting in the gap state which is not present for the $S=0$ simulation.

\begin{figure}
\begin{center}
\includegraphics[scale=0.45]{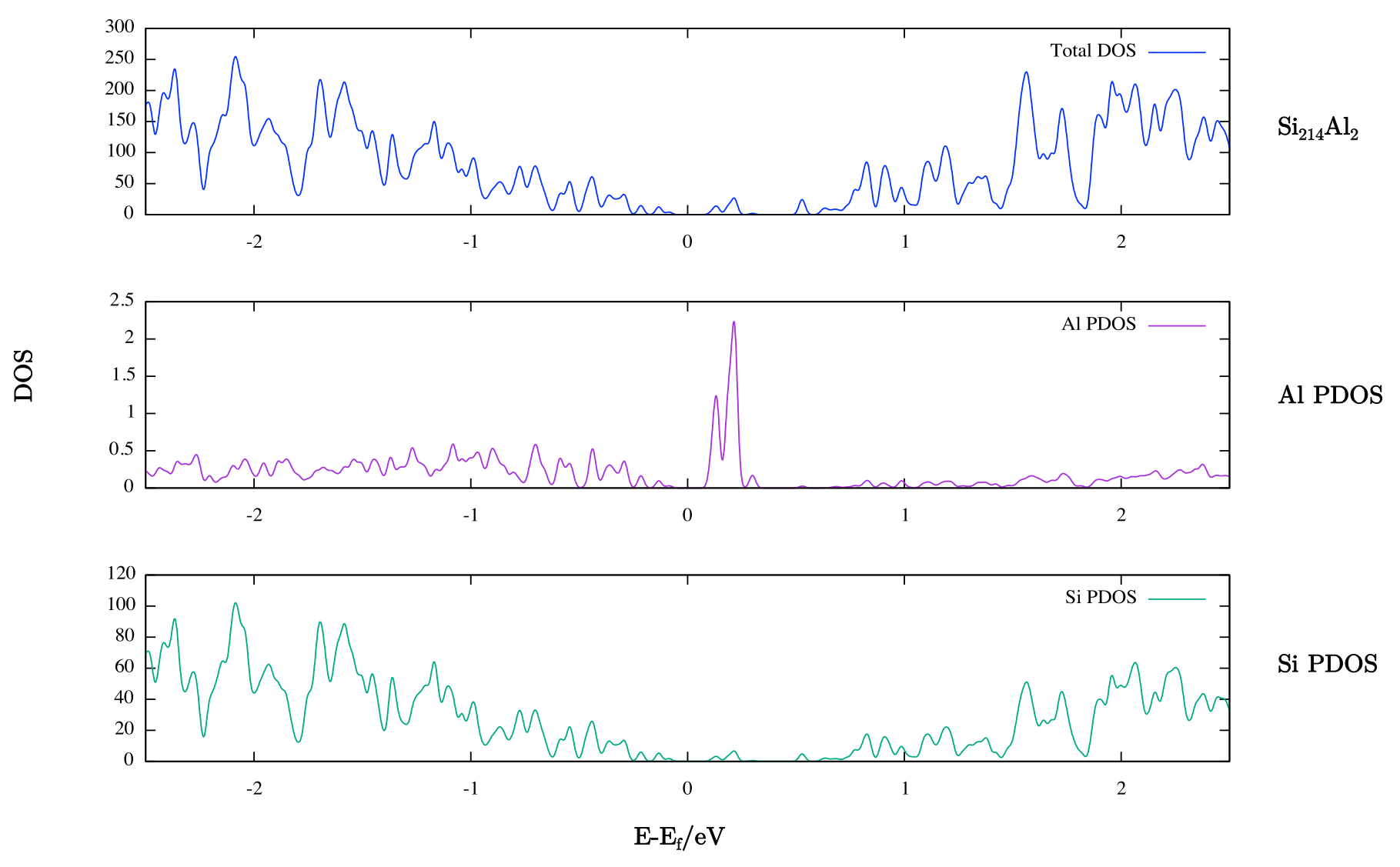}
\caption[]{The total and atom projected DOS for both undoped Si and the Si$_{214}$Al$_{2}$ structure in the $ S=0$ state  with the aluminium dopants in adjacent sites.}
\end{center}
\end{figure}

\begin{figure}
\begin{center}
\includegraphics[scale=0.45]{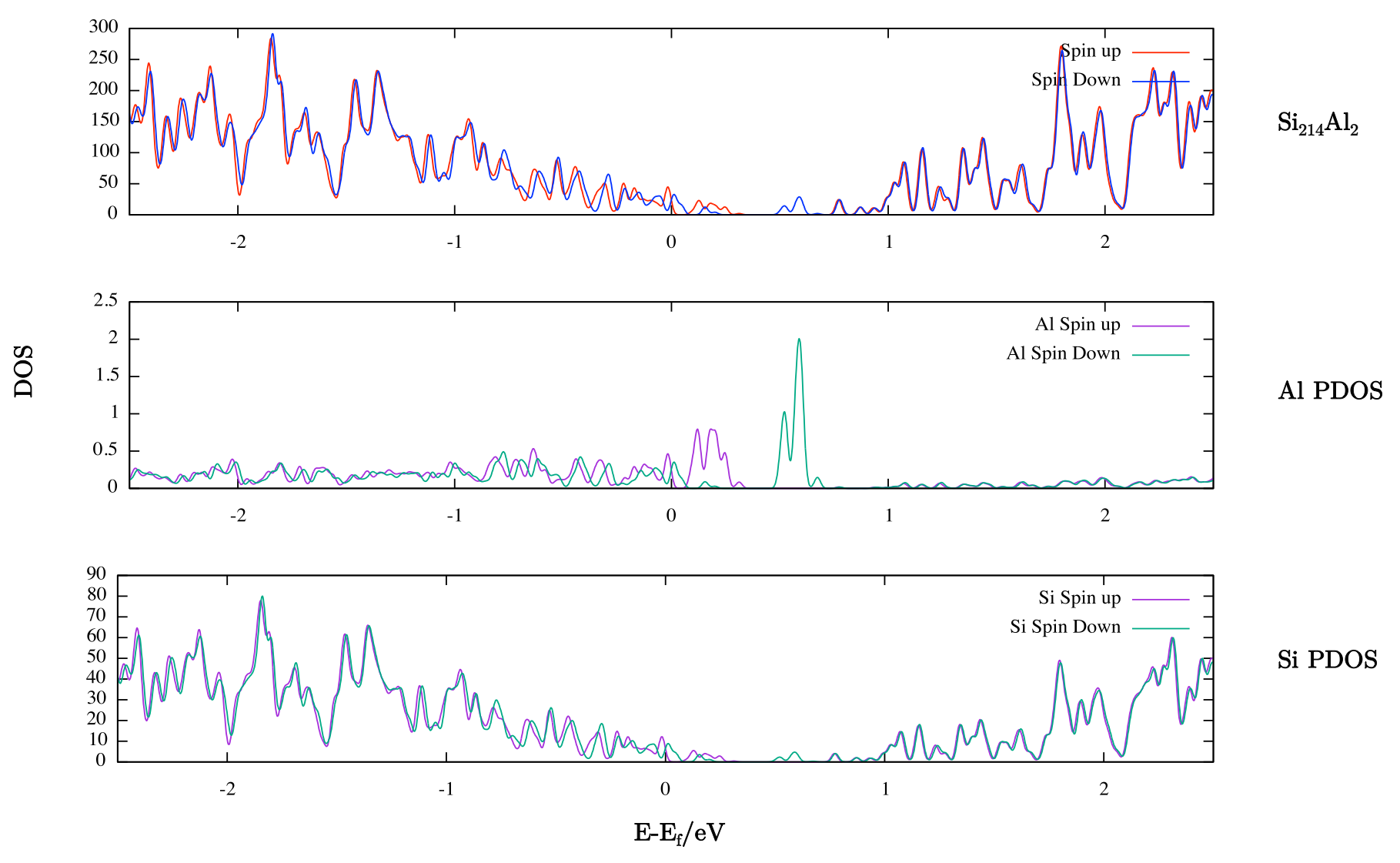}
\caption[]{The total and atom projected DOS for both undoped Si and the Si$_{214}$Al$_{2}$ structure in the $ S=1$ state  with the aluminium dopants in adjacent sites.}
\end{center}
\end{figure}

There are also differences observed between the two spin configurations of adjacent dopants when considering the VBM. The band decomposed partial charge densities for all k-points show that for the model containing adjacent dopants in the $S=1$ spin state there is a localisation of charge between the Al dopants in the highest energy occupied state, as depicted in Figure 7(b). The isosurface strongly resembles a localised dopant state. In comparison, the highest energy occupied  band for the $S=0$ spin state,  shown in Figure 7(a),  is highly delocalised.  The localisation of charge in the occupied band for the $S=1$ case combined with the high charge total density in the region between the adjacent Al dopants is suggestive of a bond. Furthermore with the majority of observable charge being more localised in the $S=1$ state than the $S=0$ state shows that the Al dopants are more likely to be active in the higher net spin configuration. 
 

\begin{figure}[h]
\hspace{1cm}
\subfloat[216 atom S=0]{
\includegraphics[scale=0.20]{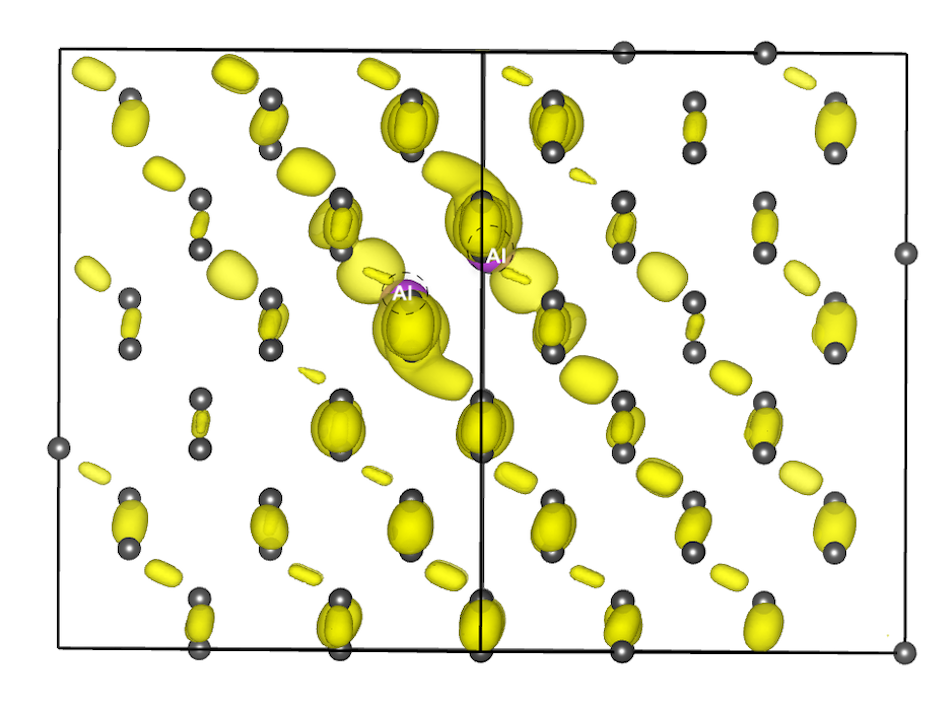}
}
\subfloat[216 atom S=1]{
\includegraphics[scale=0.20]{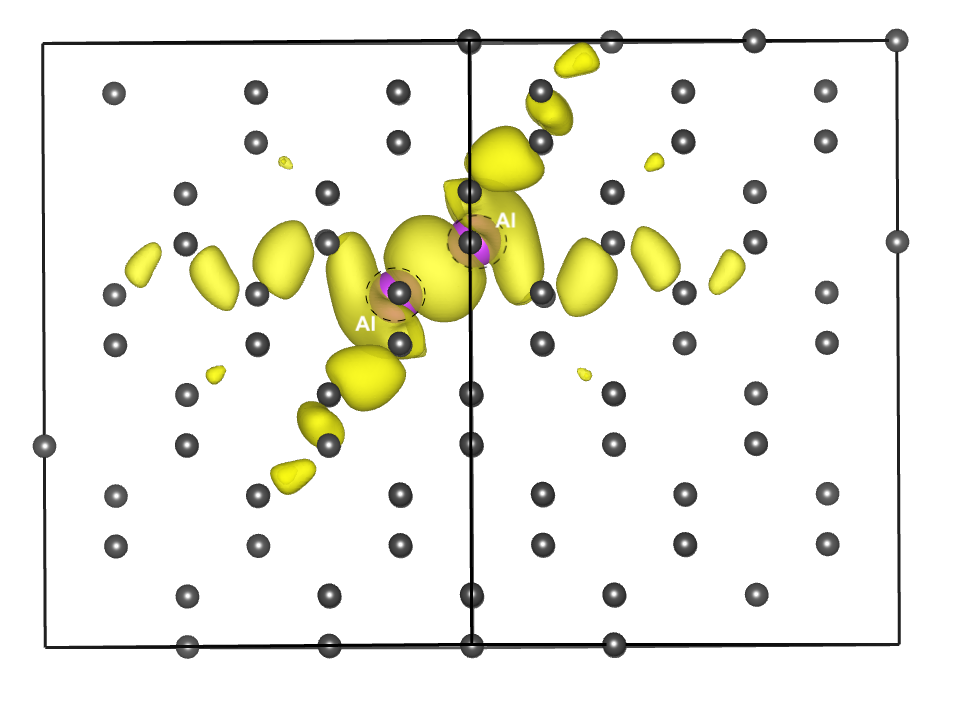}
}
\caption[]{The isosurface of the  partial charge density for the highest energy fully occupied  state  of Si$_{510}$Al$_{2}$  in the $S=0$ and the  $S=1$ spin up state with adjacent Al dopants shown in purple and labelled,  the Si is shown in gray.}
\end{figure}


\subsubsection{Electron Localisation Function for Substitutional Aluminium Pairs}\hspace*{\fill} \\

The ELF plots for both the spin polarised and non-spin polarised simulations of Al dopants in adjacent substitutional sites  are shown in Figure 8. These have been scaled to the maximum and minimum possible values of the ELF which are $0$ and $1$. As outlined in section 2.3 these values represent the probability of pairs of electrons occupying a particular region of space. While the change in bond length for the $S=0$ and the $S=1$ states may not be clear in Figure 8, the property of interest is the higher ELF value of the region between the dopants in the spin polarised simulation. The values of the ELF within this region range from $0.92$ to $0.93$,  one might associate an ELF value above 0.7 with a weakly bonded system. For the same simulation, the ELF values  between the Si-Si bonds and Al-Si bonds are approximately 0.89 and 0.88 respectively, further evidence that a bond is present between the adjacent Al dopants. For comparison simulations performed using the same convergence criteria for alane (AlH$_{3}$) and water (H$_{2}$0) had ELF values of 0.99 in the regions  where bonding would be expected to occur.  As a result we are confident that Al-Al bonds are present in the $S=1$ state, but there are no such bonds in the $S=0$ state.\\ 

\begin{figure}[h]

\vspace{-0.4cm}
\centering
\hspace{-2mm}
\subfloat[216 atom S=0]{
  \includegraphics[width=50mm]{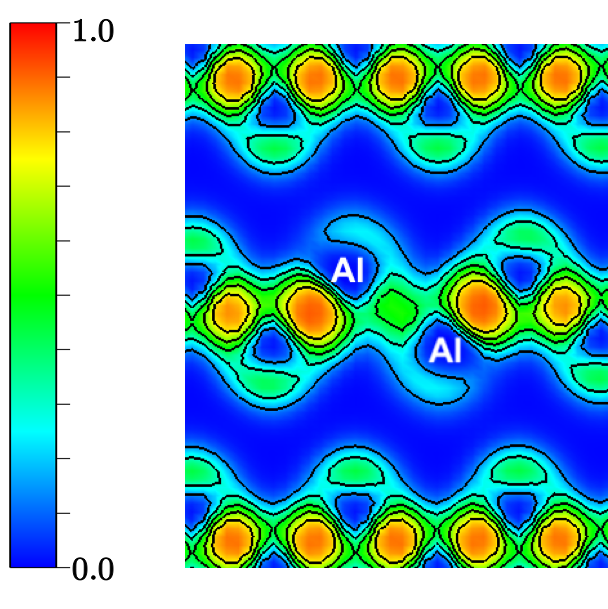}
}
\subfloat[216 atom S=2]{
  \includegraphics[width=49mm]{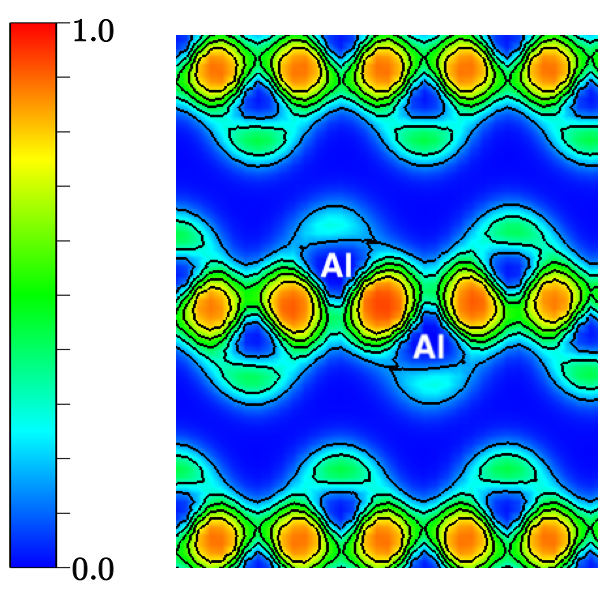}
}

\caption[]{2D ELF plots across the $\hkl{1,0,-1}$  plane containing both neighbouring Aluminium dopants in S=0 and S=1 states.}
\end{figure}

\section{Conclusions}

We have used DFT to investigate the structure and energetics of aluminium dopants within bulk silicon including multiple  dopants occupying adjacent substitutional sites. \\ 
 The interaction between the acceptor dopants has been studied using a combination of the ELF and visualisations of the densities of states. For the simple case of a single  three electron  Al dopant we observe the presence of bonds  with neighbouring Si atoms for both spin states, indicating the dopant exists in a saturated and electrically active state. \\
  For the adjacent dopants the data presented shows that a bond is formed between the Al dopants in the $S=1$ state. This is supported by a high ELF value in the region between the two acceptors, a non-negligible charge density in the same region and the separation of the Al dopants is consistent with the bond lengths cited from previous force field and DFT simulations.  However the same analysis shows that no such bond is formed between the Al dopants in the $S=0$ state. \\
 The adjacent Al dopants therefore will not be electrically active  and therefore will be self-compensating unless they occupy a high spin state in which case they will become bonded.  Hence if amine alanes are viable candidates for precision doping then aluminium dopants may be used for co-doping devices.  Furthermore, this raises the possibility of a precision doped structure which can be switched from the  $S=0$ to the $S=1$ state using a combination  optical pumping and magnetic fields,  given the low 0.11eV energy difference between the  states for the bonded Al complex. 

\section{Acknowledgments}
This work acknowledges financial support from the EPSRC ADDRFSS Programme grant (EP/M009564/1). \\ The authors are grateful for computational support from the UK Materials and Molecular Modelling Hub, which is partially funded by EPSRC (EP/P020194), for which access was obtained via the UKCP consortium and funded by EPSRC grant ref EP/P022561/1\\

\bibliographystyle{iopart-num}
\bibliography{alpaper_bibdesk}

\end{document}